\documentclass[11pt]{article}
\usepackage{jheppub}
\usepackage{braket} % bra and ket notation
\usepackage{slashed,dsfont,bm}

\usepackage[whole]{bxcjkjatype} 
\def\d{\mathrm{d}}
\newcommand{\dd}{\mathrm{d}}

\def\CB{\mathcal{B}}

\def\CK{\mathcal{K}}

\def\CM{\mathcal{M}}
\def\CN{\mathcal{N}}
\def\CO{\mathcal{O}}

\def\CQ{\mathcal{Q}}
\def\CR{\mathcal{R}}

\def\SO{\mathrm{SO}}

\title{Boundary entropy under ambient RG flow in the AdS/BCFT model}

\author{Yoshiki Sato}

\affiliation{Physics Division, National Center for Theoretical Sciences, National Tsing-Hua University, Hsinchu 30013, Taiwan}

\abstract{We discuss the change of the boundary entropy under an ambient renormalization group flow. 
We use conformal perturbation theory to calculate the change of the boundary entropy for $d$-dimensional BCFTs between two nearby fixed points.
We also use the AdS$_{d+1}$/BCFT$_d$ model to calculate the boundary entropy.
We show that the boundary entropy can increase under the ambient RG flow both in conformal perturbation theory and 
the AdS$_{d+1}$/BCFT$_d$ model.
In a special case, the change of the boundary entropy in the AdS$_{d+1}$/BCFT$_d$ model reduces to that of the conformal perturbation theory. 
}

\begin{document}
\maketitle

\section{Introduction}

The monotonicity theorems are important to understand the dynamics under a renormalization group (RG) flow in quantum field theory.
The famous monotonicity theorem is Zamolodchikov's $c$-theorem \cite{Zamolodchikov:1986gt} in two-dimensional conformal field theory (CFT).
This theorem states that there exists a $c$-function which always decreases along a RG flow triggered by a relevant operator.
In higher even dimensions, the type $A$ central charge of the conformal anomaly is conjectured to be a $c$-function \cite{Cardy:1988cwa,Myers:2010tj,Myers:2010xs} and proved later in $d=4$ \cite{Komargodski:2011vj}.
In odd dimensions, there is no conformal anomaly, but the sphere free energy is conjectured to be a $c$-function  \cite{Jafferis:2011zi,Klebanov:2011gs}.

In boundary conformal field theory (BCFT), it is known that the $g$-function decreases under a RG flow localized on the boundary \cite{Affleck:1991tk},
\begin{align}
    I = I_{\text{BCFT}} + \hat{\lambda} \int \! \dd^{d-1} \vec{x} \, \sqrt{\hat{g}}\, \hat{\CO} (\vec{x}) \,.
\end{align}
Here the coordinate system is given by $x=(\vec{x},w)$, where 
$\vec{x}$ are directions parallel to the boundary and $w$ is a direction perpendicular to the boundary at $w=0$.
The $g$-function is given by the increment of the sphere free energy due to the presence of the boundary,
\begin{align}
    \log g = Z_{\text{BCFT}}-\frac{1}{2} Z_{\text{CFT}} \,,
\end{align}
where the prefactor $1/2$ comes from the fact that the spacetime is half due to the boundary.\footnote{For defect CFTs, the boundary entropy is defined as $\log g = Z_{\text{DCFT}}-Z_{\text{CFT}}$.}
At the fixed points, this $g$-function is equal to the increment of the entanglement entropy for the spherical region,
\begin{align}
    \log g = S_{\text{BCFT}}^{\text{EE}}-\frac{1}{2} S_{\text{CFT}}^{\text{EE}} \,,
\end{align}
due to the Casini-Huerta-Myers map \cite{Casini:2011kv}.
Note that these $g$-functions are not the same along the boundary RG flow except at conformal fixed points.
The proof of the monotonicity of the $g$-function in BCFT$_2$ was first presented by \cite{Friedan:2003yc}, and then another proof based on information theory was presented by \cite{Casini:2016fgb}. 
For BCFT$_3$, the similar monotonicity theorem called $b$-theorem was proved in \cite{Jensen:2015swa} by using the similar technique in \cite{Komargodski:2011vj}. 
In higher dimensions, the  monotonicity theorem is conjectured in \cite{Nozaki:2012qd,Gaiotto:2014gha,Estes:2014hka,Kobayashi:2018lil} and the proof is presented for $d\leq 4$ \cite{Casini:2018nym} by generalizing a method in \cite{Casini:2016fgb} to higher dimensional BCFTs.\footnote{Monotonicity theorem in DCFT with higher codimensional defect has been studied in various models \cite{Jensen:2015swa,Kobayashi:2018lil,Rodgers:2018mvq}. Specifically, it is pointed out that the increment of the sphere free energy and that of the entanglement entropy in DCFT, except co-dimension one defect, are different \cite{Kobayashi:2018lil}. 
Furthermore, it is shown that the increment of the free energy is a candidate of the monotonic function while the increment of the entanglement entropy is not monotonic.}

Usually, the monotonicity theorems in BCFT or defect CFT (DCFT) are considered under the boundary or defect RG flow. 
On the other hand, the behaviour of the $g$-function under the ambient\footnote{In this paper we use the terminology `ambient' for quantities for $d$-dimensional spacetime where the $(d-1)$-dimensional boundary is embedded.
We reserve the term `bulk' for  the $(d+1)$-dimensional AdS spacetime.} RG flow,
\begin{align}
    I = I_{\text{BCFT}} + \lambda \int \! \dd^d x \, \sqrt{g} \, \CO (x) \,,
\end{align}
is not well known. 
Hence it is natural to ask whether the $g$-function is monotonic under the ambient RG flow.
This question was first addressed in \cite{Green:2007wr} and it was revealed that the $g$-function can increase under the ambient RG flow in BCFT$_2$.
To simplify calculation and discussion, 
it was assumed that the IR fixed point is close enough to the UV fixed point such that the conformal perturbation theory is valid under the RG flow triggered by only one relevant ambient operator.
The conformal dimension of the relevant operator is  slightly relevant,
\begin{align}
    \Delta = 2- \varepsilon \,, \qquad 0<\varepsilon \ll 1 \,.
\end{align}
Then the difference of the boundary entropy becomes\footnote{We use a notation different from \cite{Green:2007wr}. The coefficients $A$ and $B$ in our paper correspond to $A_\Phi^k / 2^\Delta$ and $-b$ in \cite{Green:2007wr}, respectively. 
Also, the small parameter $\varepsilon$ corresponds to $2y$ in \cite{Green:2007wr} for $d=2$.} 
\begin{align} \label{change}
    \frac{\delta g}{g} = \frac{2A}{B}\varepsilon + \CO (\varepsilon^2) \,, 
\end{align}
where $A$ is a coefficient of the one-point function,
\begin{align}
    \langle \CO (\vec{x},w) \rangle = \frac{A}{ w^{\Delta}} \,,
\end{align}
and $B$ is a coefficient of three-point function in CFT without the boundary,
\begin{align} \label{three-point}
    \langle \CO (x_1) \CO (x_2) \CO (x_3) \rangle_{\text{CFT}} = \frac{B}{ |x_1-x_2|^\Delta |x_2-x_3|^\Delta |x_3-x_1|^\Delta } \,.
\end{align}
The equation \eqref{change} shows that the change of the boundary entropy depends on the relative signs of $A$ and $B$, and the boundary entropy can increase.

The main subject of this paper is to study the change of the boundary entropy in the holographic dual of BCFT  proposed by Takayanagi \cite{Takayanagi:2011zk,Fujita:2011fp}.
The entanglement entropy in BCFT is a useful quantity to prove properties of the boundaries.
Furthermore, it is recently revealed that the entanglement entropy in BCFT and the holographic entanglement entropy \cite{Ryu:2006bv} in the AdS/BCFT model are important for solving the  information paradox of black holes \cite{Almheiri:2019hni}. In this situation, we study the entanglement entropy under the ambient RG flow in the AdS/BCFT model and provide  evidence for the non-monotonicity of the $g$-function as an application.
In the AdS/BCFT model, it is possible to compute the change of the boundary entropy without assuming that the relevant operator is slightly relevant.
It is also easy to work in higher dimensions.
We can realise the slightly relevant operator deformation in the the AdS/BCFT model as a special case.
To compare the holographic result with CFT, we also work in conformal perturbation theory in BCFT$_d$.

The organization of this paper is as follows.
In section \ref{sec_cpt}, we calculate the change of the boundary entropy using conformal perturbation theory in CFT$_d$.
In section \ref{sec2}, we review the AdS/BCFT model and discuss a domain wall flow which is a dual description of the ambient RG flow.
In section \ref{sec3}, we concentrate on the holographic description of the RG flow by the slightly relevant operator, and we check whether our holographic calculation is consistent with conformal perturbation theory in section \ref{sec_cpt}.
The final section is devoted to discussion.

\section{Conformal perturbation theory}
\label{sec_cpt}

In this section, we calculate the change of the boundary entropy using conformal perturbation theory in BCFT$_d$.
This is a higher dimensional generalization of the work \cite{Green:2007wr}, but we use a different coordinate system and a regularization method even in BCFT$_2$.

We locate BCFT$_d$ on a hemisphere with radius $R$ whose metric is 
\begin{align}
    \dd s^2 = R^2 (\dd \theta^2 + \sin^2 \theta \dd s_{\mathbb{S}^{d-1}}^2) \,,
\end{align}
where the range of $\theta$ is $0\leq \theta \leq \pi/2$ and the boundary sits at $\theta = \pi/2$.
We perturb it by an ambient RG flow,
\begin{align}
    I = I_{\text{BCFT}} + \lambda \int \! \dd^d x \, \sqrt{g}\, \CO (x) \,, 
\end{align}
and we are interested in the change of the boundary entropy under the RG flow.

Now we consider the case where the operator $\CO$ is a slightly relevant operator with a conformal dimension,
\begin{align}
    \Delta = d -\varepsilon \,,
\end{align}
where $\varepsilon$ is very small, $0<\varepsilon \ll 1$.
The theory flows to a nontrivial fixed point at 
\begin{align} \label{lambda_fixed}
    \lambda \mu^{-\varepsilon}= \frac{2}{\text{Vol}(\mathbb{S}^{d-1}) B} \varepsilon
\end{align}
for positive $B$ \cite{Klebanov:2011gs}.\footnote{In \cite{Klebanov:2011gs}, the conformal coupling \eqref{lambda_fixed} is obtained for CFT$_d$ without boundary. When the boundary is introduced, the conformal coupling \eqref{lambda_fixed} does not change since the boundary does not change the ambient theory.}
Here the parameter $\mu$ is the renormalization scale, and it is related with the radius $R$ through $\mu = 1/(2R)$.
$B$ is the coefficient of the three-point function \eqref{three-point}, and  
 $\text{Vol}(\mathbb{S}^{d-1}) = 2\pi^{d/2}/\Gamma (d/2)$ is a volume of a $(d-1)$-dimensional sphere.
By using conformal perturbation theory, the change of the free energy becomes,
\begin{align}
    \delta Z &= -\lambda \int \! \dd^d x \, \sqrt{g} \, \langle \CO (x) \rangle \\
    & = - \lambda  R^{\varepsilon} A \text{Vol}(\mathbb{S}^{d-1}) \frac{\Gamma (d/2) \Gamma((1-d+\varepsilon)/2) }{2 \Gamma ((1+\varepsilon)/2)} \,, 
\end{align}
where we use the expression of the one-point function 
\begin{align}
    \langle \CO (x ) \rangle = \frac{A}{(R\cos \theta)^\Delta}
\end{align}
and a dimensional regularization to perform the $\theta$ integral.
In total, the change of the free energy is given by
\begin{align}
    \delta Z =  \frac{\pi^{1/2} \Gamma (d/2) }{\sin (\pi(d-1)/2) \Gamma ((d+1)/2) } \cdot \frac{A}{B}\varepsilon \,,
    \label{change_free}
\end{align}
due to the fact that the coupling constant  \eqref{lambda_fixed} of the IR fixed point is order  $\CO(\varepsilon)$. 
The following identity is also used:
\begin{align}
    \Gamma \left( \frac{1-d}{2}  \right) = \frac{\pi}{\sin (\pi(1-d)/2) \Gamma ((d+1)/2) } = - \frac{\pi}{\sin (\pi(d-1)/2) \Gamma ((d+1)/2) } \,.
\end{align}
For $d=2$, our result \eqref{change_free} is the same as that of \cite{Green:2007wr}.
Note that our coordinate system and a regularization method are different from those of \cite{Green:2007wr} but we obtain the same result.

For a higher dimensional case, we have to extract a universal part of the boundary entropy, which does not depend on a regularization scheme, from \eqref{change_free}.
Since we are working in dimensional regularization, the universal part of the boundary free energy can be obtained as 
\begin{align} \label{univ_free_energy}
    D = \sin \left(\frac{\pi(d-1)}{2} \right) \delta Z = \frac{\pi^{1/2} \Gamma (d/2) }{\Gamma ((d+1)/2) } \cdot \frac{A}{B}\varepsilon \,,
\end{align}
as explained in \cite{Giombi:2014xxa}.
As in $d=2$ case, both the change of the free energy \eqref{change_free} and its  universal part \eqref{univ_free_energy} are proportional to the ration of the one-point function and the three-point function, and the ration does not have a definite sign.
Then, the boundary entropy can increase under the ambient RG flow in BCFT$_d$.

\section{AdS/BCFT model}
\label{sec2}

The AdS/BCFT model is a bottom-up model of a holographic dual of BCFT introduced by Takayanagi \cite{Takayanagi:2011zk,Fujita:2011fp}.
In this section, we review the AdS/BCFT model briefly and discuss the RG flow triggered by a single relevant operator.
We also discuss a general behaviour of the boundary entropy under the RG flow.

We construct the gravity dual following Takayanagi's proposal
by introducing the AdS boundary $\CQ$ with a brane of tension $T$.
The gravity action in the Euclidean signature is given by 
\begin{align}
	I = - \frac{1}{16\pi G_\text{N}} \int_{\CB}\sqrt{G}\, \left( \CR + \frac{d(d-1)}{L^2}\right)  - \frac{1}{8\pi G_\text{N}} \int_{\CQ}\sqrt{\hat{G}} \left( \CK - T\right) - \frac{1}{8\pi G_\text{N}} \int_{\CM}\sqrt{\hat{G}}\, \CK \, ,
\end{align}
where $\CB$ is the bulk AdS space and $\CM$ is the boundary on which the dual BCFT lives.
The first term is the Einstein-Hilbert action with the bulk metric $G_{MN}$ and the AdS radius $L$.
To make the variational problem well-defined in the presence of the boundary, the Gibbons-Hawking terms are introduced with the extrinsic curvature defined by 
\begin{align}
	\CK_{MN} = \hat{G}_{ML}\hat{G}_{NK}\nabla^L n^K
\end{align}
for the outward pointing normal vector $n^M$.
Here $\hat{G}_{MN}$ represents the induced metric on $\CQ$ or $\CM$.
The Dirichlet boundary condition is imposed on $\CM$, while the Neumann boundary condition is imposed on $\CQ$:
\begin{align} \label{bc_q}
	\CK_{MN} -\hat{G}_{MN}\,\CK = - T\, \hat{G}_{MN} \,.
\end{align}

Now we take the ansatz for the metric such that the metric is a pure AdS metric, 
\begin{align}\label{BCFT_coord}
	\d s^2 = L^2 \frac{\dd z^2 +\eta_{\mu \nu} \dd x^\mu \dd x^\nu}{z^2} \,,
\end{align}
where $\mu , \nu=1,\cdots ,d$.
From the boundary condition \eqref{bc_q}, the shape of the brane $\CQ$ is determined as $z= x_1 /v$ where the slope parameter $v$ is related with the tension $T$.
Then the presence of the brane $\CQ$ reduces the isometry of AdS from $\SO (1,d+1)$ to $\SO (1,d)$, which is the isometry of BCFT$_d$.
It is convenient to use the following coordinate system:
\begin{align}\label{BCFT_coord_2}
	\d s^2 =   \dd \rho^2 + L^2 \cosh^2\left(\frac{\rho}{L}  \right) \frac{\dd w^2 + \delta_{ij} \dd x^i \dd x^j }{w^2} \,,
\end{align}
where $i,j = 2,\cdots ,d$.
This metric \eqref{BCFT_coord_2} can be obtained by changing the variables 
\begin{align}
    z = \frac{w}{\cosh (\rho / L)} \,, \qquad 
    x_1 = w \tanh \left( \frac{\rho}{L} \right) \,.
\end{align}
The boundary, $z= x_1 /v $, is mapped to $\rho = \rho_0$ with $\sinh (\rho_0/L) = v$.
In the new coordinate system \eqref{BCFT_coord_2}, the isometry of BCFT$_d$ is explicit since the metric  contains the metric of AdS$_d$.
Now the range of $\rho$ is $-\infty < \rho < \rho_0$, where $\rho = - \infty$ corresponds to the boundary $\CM$ where the dual BCFT lives, and $\rho_0$ corresponds to a position of the brane $\CQ$.
Since the extrinsic curvature is given by
\begin{align}
	\CK = \frac{d}{L}\,\tanh \left( \frac{\rho}{L} \right) \, ,
\end{align}
for any constant $\rho$ surface, the brane tension is fixed to be
\begin{align}
\label{tension}
	T = \frac{d-1}{L}\,\tanh \left( \frac{\rho_0}{L} \right) \, .
\end{align}
In the AdS/BCFT model, the boundary entropy can be obtained by evaluating the on-shell action.
The on-shell action is given by 
\begin{align}
\begin{aligned}
    I (\rho_0) = \frac{L^{d-1}}{8\pi G_{\text{N}}}\text{Vol}(\mathbb{H}^d) & \left[ \frac{d}{L} \int_{-\infty}^{\rho_0} \dd \rho \, \cosh^d \left( \frac{\rho}{L}\right) -\frac{TL}{d-1}\cosh^d \left( \frac{\rho_0}{L}\right) \right. \\
    & \qquad \left. +d \lim_{\rho \to -\infty} \tanh \left( \frac{\rho}{L}\right)\cosh^d \left( \frac{\rho}{L}\right)  \right]
    \end{aligned}
\end{align}
without regularization.
Then the boundary entropy is the difference of the free energy and it is given by
\begin{align} 
    \log g & = - I (\rho_0) + I(0) \\
    &= \frac{L^{d-1}}{4G_{\text{N}}} \frac{\pi^{(d-1)/2}}{\sin (\pi (d-1)/2) \Gamma ((d-1)/2) } \tanh \left( \frac{\rho_0}{L}\right)
    {}_2F_1 \left( \frac{1}{2},\frac{d}{2},\frac{3}{2}, \tanh^2 \left( \frac{\rho_0}{L}\right) \right) \,,
    \label{g-func-general}
\end{align}
where we use a regularized expression for the volume of hyperbolic space,
\begin{align}
    \text{Vol}(\mathbb{H}^d) = \frac{\pi^{(d+1)/2}}{\sin (\pi (d+1)/2)\Gamma ((d+1)/2)} \,.
\end{align}
In particular, the boundary entropy has a simple expression,  
\begin{align} \label{g-func}
    \log g = \frac{\rho_0}{4G_{\text{N}}} \,, 
\end{align}
for $d=2$.
It is shown that this $g$-function \eqref{g-func} decreases along the RG flow triggered by a relevant operator localized on boundary \cite{Takayanagi:2011zk,Fujita:2011fp,Nozaki:2012qd,Kobayashi:2018lil}.
This can be understood as a decreasing of the position of the brane under the boundary RG flow.
In this paper, we address the question of which direction the position of the brane moves by the ambient RG flow instead of the boundary RG flow. 

Let us move to a discussion about a holographic description of an ambient RG flow triggered by a relevant operator $\mathcal{O}$ with a conformal dimension $\Delta (<d)$.
Let $\phi$ be a dual scalar field of $\mathcal{O}$.
The ambient RG flow is usually realised as the domain wall flows.
Without the boundary $\CQ$, the action of the scalar field in Euclidean signature is given by 
\begin{align}
     I_\phi = \int \! \dd^{d+1}X \, \sqrt{G} \left( \frac{1}{2} (\partial \phi) ^2  +V(\phi) \right)  
\end{align}
where the potential $V(\phi)$ includes a mass term $m^2 \phi^2 /2$ and the mass is related with the conformal dimension $\Delta$ as $m^2 L^2 =\Delta (\Delta - d)$.
It is assumed that the potential has two extrema at least. The RG flow corresponds to a flow from an extremum at $\phi_{\text{UV}}$ to another extremum at $\phi_{\text{IR}}$ such that 
\begin{align}
    V(\phi_{\text{UV}}) > V(\phi_{\text{IR}}) \,.
\end{align}
Hence, the AdS radius changes from $L$ to $L_{\text{IR}} <L$ with 
\begin{align}
    \frac{d(d-1)}{L^2} - 16 \pi G_{\text{N}} V(\phi_{\text{IR}})= \frac{d(d-1)}{L_{\text{IR}}^2}
\end{align}
when $\phi_{\text{UV}}=0$ and $V(\phi_{\text{UV}})=0$.

Next, we introduce the effect of the presence of the boundary.
This is achieved by introducing the interaction between the scalar field and the boundary, 
\begin{align}
    I_{\text{int}} = \int_{\CQ} \! \dd^d X \, \sqrt{\hat{G}} V_\text{int} (\phi) \,.
\end{align}
At the IR fixed point, the brane $\CQ$ has an effective brane tension $T + 8 \pi G_{\text{N}} V_{\text{int}} ( \phi_{\text{IR}})$.
Then the position of the brane $\CQ$ is determined as 
\begin{align} \label{change_gen}
    T + 8 \pi G_{\text{N}} V_{\text{int}} ( \phi_{\text{IR}}) = \frac{d-1}{L_{\text{IR}}} \tanh \left( \frac{\rho_{\text{IR}}}{L_{\text{IR}}} \right) \,.
\end{align}
Without the interaction term, $\rho_{\text{IR}}$ is always smaller than $\rho_{0}$ since the AdS radius becomes smaller under the RG flow.
However, $\rho_{\text{IR}}$ can be larger than $\rho_0$ due to the interaction.
From \eqref{tension} and  \eqref{change_gen}, the potential $V_{\text{int}} ( \phi_{\text{IR}})$  can be written as 
\begin{align}
\label{318}
    \frac{8 \pi G_{\text{N}}}{d-1}  V_{\text{int}} ( \phi_{\text{IR}}) = \frac{1}{L_{\text{IR}}} \tanh \left( \frac{\rho_{\text{IR}}}{L_{\text{IR}}} \right) - \frac{1}{L}\,\tanh \left( \frac{\rho_0}{L} \right) \,.
\end{align}
If $\rho_{\text{IR}}<\rho_0$, the right hand side of \eqref{318} is bounded by $\tanh ( \rho_0/{L_{\text{IR}}} ) / L_{\text{IR}} - \tanh (\rho_0/L) /L$, and this function has a maximum.
Hence, $\rho_{\text{IR}}$ should be larger than $\rho_0$ when the potential is larger than the maximum.
This fact shows that the boundary entropy can increase along the ambient RG flow since the boundary entropy \eqref{g-func-general} is a monotonic function of $\rho_0$.

\section{Slightly relevant ambient RG flow in the AdS/BCFT model}
\label{sec3}

In the previous section, we discussed the boundary entropy in the general RG flow in the AdS$_{d+1}$/BCFT$_d$ model.
In this section, we restrict our attention to a situation where the  operator is slightly relevant and conformal perturbation theory is valid along the RG flow.
In $d=2$, our setup reduces to a holographic description of \cite{Green:2007wr}, and we derive the change of the boundary entropy \eqref{change} holographically.

Let us consider the dual scalar field has the following action:
\begin{align}
    I_\phi = \int \! \dd^{d+1}X \sqrt{G} \left( \frac{1}{2}G^{MN}\partial_M \phi \partial_N \phi +\frac{1}{2}m^2 \phi^2 +\frac{b}{3!} \phi^3 \right) \,,
\end{align}
with the metric \eqref{BCFT_coord_2} and the mass $m^2 L^2 = \Delta (\Delta-d)$.
Here, the fact that the potential term only includes up to $\phi^3$ means that we consider only a leading contribution of conformal perturbation theory.
In the AdS$_{d+1}$/CFT$_d$ correspondence, the two-point function is given by 
\begin{align} \label{2pt}
    \langle \CO (x_1) \CO (x_2) \rangle_{\text{CFT}} = L^{d-1} \frac{(2\Delta -d)\Gamma (\Delta)}{\pi^{d/2} \Gamma (\Delta -d/2)} \frac{1 }{|x_1-x_2|^{2\Delta}} \,,
\end{align}
and the three-point function is given by 
\begin{align}
    \langle \CO (x_1) \CO (x_2) \CO (x_3)\rangle_{\text{CFT}} = b L^{2} \frac{\Gamma (\Delta/2)^3\Gamma((3\Delta -d)/2)}{2\pi^d \Gamma (\Delta -d/2)^3} \frac{1}{|x_1-x_2|^\Delta |x_2-x_3|^\Delta |x_3-x_1|^\Delta} \,.
\end{align}
Note that these correlation functions are those of CFT without boundary and we add the subscript `CFT' in the above expressions to distinguish correlation functions in BCFT. 
Since the two-point function is not canonically normalized, we have to normalize the operator $\CO$ when we compare our holographic results with the CFT results.

Next, we consider the contribution of the boundary $\CQ$.
We assume the interaction between the scalar field and the brane as 
\begin{align}
    V_\text{int} (\phi ) = a \phi \,.
\end{align}
Here the parameter $a$ is related with a coefficient of the one-point function, 
\begin{align}
\begin{aligned}
    \langle \CO (x) \rangle & = \frac{aL \pi (2\Delta -d) \sin (\pi \Delta/2) \Gamma (\Delta)}{w^\Delta 2^{\Delta} \sin (\pi (\Delta-d/2)) \Gamma (\Delta-d/2+1) \Gamma ((d-\Delta+1)/2) \Gamma ((\Delta+1)/2)} \\
    & \quad \times \frac{1}{\sqrt{1+v^2} {}_2 F_1 ((\Delta+1)/2 ,(d-\Delta+1)/2 , 1/2, -v^2 ) }\,,
    \end{aligned}
\end{align}
with $v= \sinh (\rho_0/L)$ as before.
Recall that $w$ is the direction perpendicular to the boundary.
See section 8 and appendix D in \cite{Fujita:2011fp} for the details of the computation of the one-point function.
After normalizing the two-point function \eqref{2pt} canonically, the coefficients of the one-point function in BCFT and the three-point function in CFT become
\begin{align}
\begin{split}
    A & = \frac{aL \pi (2\Delta -d )\sin (\pi \Delta/2) \Gamma (\Delta)}{\CN 2^{\Delta} \sin (\pi (\Delta-d/2)) \Gamma (\Delta-d/2+1) \Gamma ((d-\Delta+1)/2) \Gamma ((\Delta+1)/2)} \\
    & \quad \times \frac{1}{\sqrt{1+v^2} {}_2 F_1 ((\Delta+1)/2 ,(d-\Delta+1)/2 , 1/2, -v^2 ) } \,,
    \end{split} \\
    B & = \frac{bL^2}{\CN^3}  \frac{\Gamma (\Delta/2)^3\Gamma((3\Delta -d)/2)}{2\pi^{d} \Gamma (\Delta -d/2)^{3}} \,,
\end{align}
respectively. 
Here $\CN$ is a normalization constant,
\begin{align}
    \CN^2 = L^{d-1} \frac{(2\Delta-d)\Gamma (\Delta) }{\pi^{d/2}\Gamma (\Delta -d/2)} \,,
\end{align}
for the scalar primary operator to be canonically normalized.
The bulk potential $V(\phi) = m^2 \phi^2 / 2 + b \phi^3 /3! $ has two extrema: $\phi=0$ and $\phi = -2m^2 / b$.
The first extrema $\phi_{\text{UV}}=0$ corresponds to the UV fixed point, and the second extrema $\phi_{\text{IR}} = -2 m^2 /b$ corresponds to the IR fixed point.

Let us consider the situation where the conformal dimension is slightly relevant:
\begin{align}
    \Delta = d - \varepsilon \,, \qquad 0 < \varepsilon \ll 1 \,.
\end{align}
Then, the equation \eqref{change_gen} reduces to 
\begin{align}
    T + 8 \pi G_{\text{N}} a \phi_{\text{IR}} = \frac{d-1}{L} \tanh \left( \frac{\rho_0 + \delta \rho}{L} \right) \,,
\end{align}
where we use the fact that the change of the AdS radius can be ignored because it is order $\CO (\varepsilon^2)$.
From this relation, the change of the position of the brane becomes,
\begin{align} \label{4.11}
    \delta \rho = \frac{8\pi G_{\text{N}}}{d-1} a L^2 \phi_{\text{IR}} \cosh^2 \left( \frac{\rho_0}{L} \right) \,.
\end{align}
From \eqref{g-func-general} and \eqref{4.11}, the change of the boundary entropy becomes 
\begin{align}
    \delta \log g 
    = \frac{2\pi }{d-1} a \phi_{\text{IR}} \left( L \cosh \left( \frac{\rho_0}{L} \right)  \right)^{d} \frac{\pi^{(d-1)/2}}{\sin (\pi (d-1)/2) \Gamma ((d-1)/2) } \,.
\end{align}
Furthermore, by using the ratio between the coefficient of the normalized one-point function in BCFT and that of the three-point function in CFT,  
\begin{align}
     \frac{A}{B} = \frac{2d a \pi^{d/2}}{b\Gamma (d/2)} L^{d-2} \left( \cosh \left( \frac{\rho_0}{L}\right) \right)^d \,,
\end{align}
and the approximated mass,
\begin{align}
    m^2 L^2 \simeq - d \varepsilon \,,
\end{align}
the change of the boundary entropy becomes
\begin{align} \label{change_hol}
    \delta \log g 
    = \frac{\pi^{1/2}\Gamma (d/2)}{\sin (\pi (d-1)/2) \Gamma ((d+1)/2) } \frac{A}{B} \varepsilon \,.
\end{align}
This is exactly the same result of the conformal perturbation theory computation in section \ref{sec_cpt} and our analysis is consistent with conformal perturbation theory.
The universal part can be extracted as
\begin{align} \label{change_hol_univ}
    \delta \log g |_{\text{univ}} =  \frac{\pi^{1/2}\Gamma (d/2)}{\Gamma ((d+1)/2) } \frac{A}{B} \varepsilon \,,
\end{align}
and this is also the same as the universal part of the boundary entropy obtained by conformal perturbation theory \eqref{univ_free_energy}.

\section{Discussion}

In this paper, we discussed the change of the boundary entropy under the ambient RG flow instead of the boundary RG flow both in conformal perturbation theory in BCFT$_d$ and the AdS$_{d+1}$/BCFT$_d$ model.
We showed that the change of the boundary entropy obtained by conformal perturbation theory is proportional to the ratio of the coefficient of the one-point function in BCFT and the coefficient of the three-point function in CFT. Then the boundary entropy does not always decrease along the ambient RG flow.
We also showed that the change of the boundary entropy obtained from \eqref{change_gen} depends on the bulk potential and the interaction between the bulk scalar and the brane $\CQ$, and hence the boundary entropy does not necessary decrease along the ambient RG flow. 
Specifically, for a slightly relevant ambient perturbation case, the change of the boundary entropy CFT \eqref{change_hol} is the same as that of conformal perturbation theory \eqref{change_free}.

It is natural to expect the existence of the monotonic function under the ambient RG flow since the ambient RG flow connects two BCFTs.
Even if there is a boundary, the  $c$-function associated with the ambient CFT decreases under the ambient RG flow.
However, this $c$-function does not contain information of the presence of the boundary.
Hence, a natural candidate is the $g$-function, or the boundary entropy, but this paper shows that the $g$-function is not a monotonic function.

There are several future directions.
In this paper we only consider a slightly relevant deformation as a holographic example. However, an exact marginal deformation has also been discussed in \cite{Green:2007wr} (See also recent works \cite{Herzog:2019rke,Bianchi:2019umv} for marginal deformations in BCFT and DCFT).
It is interesting to consider the exact marginal deformation in the AdS/BCFT model.

Recently, BCFT$_3$ and DCFT with a two-dimensional defect are studied in various papers \cite{Jensen:2015swa,Rodgers:2018mvq,Estes:2018tnu,Jensen:2018rxu,Chalabi:2020iie}. 
In BCFT$_3$ and DCFT with a two-dimensional defect, it is known that $b$-function, which is a coefficient of trace anomaly, decreases along boundary or defect RG flow \cite{Jensen:2015swa}.
In BCFT$_3$, we revealed that the boundary entropy can increase under the ambient RG flow.
Hence, it is interesting to ask how the $b$-function in DCFT changes along the ambient RG flow.

\acknowledgments

The author would like to thank C.~S.~Chu and S.~Kawamoto for discussion.
This work is supported by the National Center of Theoretical Sciences (NCTS).

\bibliographystyle{JHEP}
\bibliography{Defect_Entropy}

\end{document}